\documentstyle[preprint,aps]{revtex}

\tighten
\input epsf

\begin{document}
\preprint
{NSF-ITP-95-58 \quad gr-qc/9506078 \quad {\sl Phys.~Rev.~D}, in press}

\title{Critical Exponents and Stability at the Black Hole Threshold
	for a Complex Scalar Field}
\author{Eric W. Hirschmann\footnote%
	{Electronic address: \tt ehirsch@dolphin.physics.ucsb.edu\hfil}}
\address{Department of Physics\\
	University of California\\
	Santa Barbara, CA 93106-9530}
\author{Douglas M. Eardley\footnote%
	{Electronic address: \tt doug@itp.ucsb.edu\hfil}}
\address{Institute for Theoretical Physics\\
	University of California\\
	Santa Barbara, CA 93106-4030}
\date{\today}
\maketitle

\begin{abstract}
This paper continues a study on Choptuik scaling
in gravitational collapse of a massless complex scalar field at the
threshold for black hole formation.  We perform a linear
perturbation analysis of the previously derived complex critical
solution, and calculate the critical exponent for black hole mass,
$\gamma \approx 0.387111\pm0.000003$.  We also show that this critical
solution is unstable via a growing oscillatory mode.
\end{abstract}

\narrowtext

\pacs{04.20.Jb}

\section{Introduction}

The recent discovery by Choptuik and others \cite{Chop,AE} of new
and unexpected critical behavior in gravitational collapse has
sparked efforts, both numerical and analytic, to understand these
phenomena.  One would especially like to understand the nature of
the black hole scaling relation, and how the observed echoing
arises out of the Einstein equations.  That these phenomena appear
for a variety of different matter fields suggests that they may
reflect universal properties of the Einstein equations.  Choptuik
originally conjectured that the exponent, $\gamma$, in the
threshold scaling relation for black holes mass,
\begin{equation}
M_{\rm BH}(p) \propto \cases{
	0,						&$p\le p^*$\cr
	(p-p^*)^\gamma \quad (\gamma\approx0.37 ),	&$p>p^*$\cr}
\label{bhmass}\end{equation}
was universal for collapse of a scalar field, and presented
considerable evidence.  Here $p$ is some parameter which
characterizes the strength of the initial conditions, and $p^{*}$
is the threshold value, {\it i.e.,} the value for the critical
solution.  Work by Abrahams and Evans \cite{AE} on vacuum collapse
in axial symmetry, and Evans and Coleman \cite{EC} on collapse of a
radiation fluid, suggested a broader universality or
near-universality of $\gamma$ (see also \cite{KHA}) among different
matter models.  All these results were obtained by numerical
relativity, so that accuracy of $\gamma$ was at best $10^{-3}$.
However, Maison \cite{Maison} showed that fluid collapse models
with an equation of state given by $p = k\rho$ have a critical
exponent $\gamma$ that depends strongly on the parameter $k$ in the
range $0\le k\le0.88$\@.  (Evans and Coleman had examined only the
case for radiation fluid, $k = 1/3$.) Maison used an accurate
perturbation theory method, as suggested by Evans and Coleman, and
carried out by Koike, Hara \& Adachi \cite{KHA}.  Perturbation
theory affords both a calculation of $\gamma$, and a test of
stability.

Just at the critical value $p=p^*$ --- at the ``black hole
threshold' --- a critical solution of the field equation appears, a
``choptuon".  A choptuon is a delicately poised dynamical state of
the fields, a collapsing, shrinking, radiating ball of field
energy.  Its self-gravity gives rise to spacetime curvature that
partially traps field energy, but only partially:  Energy
continually leaks outward and becomes outgoing radiation.  By
careful tuning of $p$, the rate of collapse exactly balances the
radiation rate, so that the collapse becomes self-similar, and the
choptuon shrinks all the way down to the Planck scale.  This
situation is of course intrinsically unstable:  If self-gravity is
a little too weak, the choptuon will soon dissipate completely;  if
a little too strong, it will soon form an event horizon and undergo
terminal gravitational collapse into a black hole much larger than
the Planck mass.  Construction of the choptuon implies
stabilization, by careful tuning of a parameter $p$, against this
threshold instability toward either subcritical dissipation or
supercritical black hole formation.  We will call this instability
the ``black hole instability" for short, bearing in mind that its
other side is dissipation.  The critical exponent $\gamma$ measures
the strength of the black hole instability.

Then the question of stability of the choptuon is this:  Does
the choptuon have any {\it other} instabilities, aside from the
inevitable black hole instability?  The previous evidence is
that choptuons are in fact stable in this sense \cite{Chop,AE}.

In this paper we return to the gravitational collapse of a massless
complex scalar field in spherical symmetry, as a good theoretical
laboratory in which to address these issues.  In \cite{HE} ---
hereinafter called paper I --- we derived a new continuously
self-similar choptuon at the threshold of black hole formation, and
constructed the complete spacetime.  The discrete echoing seen in
the original choptuon \cite{Chop} reappeared as continuous phase
oscillations of the complex scalar field.  However, the choptuon by
itself does not yield the black hole scaling relation; one needs to
perturb the choptuon.  In addition, perturbation theory can probe
important questions about the stability of choptuons, and the
nature of the relative attractor in the space of all initial
conditions.  This paper considers only spherically symmetric
perturbations, the relevant ones for Eq.~(\ref{bhmass}), and the
most likely ones to be unstable.  It would also be interesting to
carry out a nonspherical perturbation analysis of any of the known
choptuons, to check Choptuik's ``no hair" conjecture \cite{Chop}.

The properties of the critical solution are described in
paper I, and will not be repeated here.  Since there is confusion
in the literature on this point, we emphasize again that the
self-similar hypothesis --- whether discrete \cite{Chop} or
continuous \cite{EC,HE,EHH} --- is {\it not} unphysical for
gravitational collapse in asymptotically flat spacetime.  No
self-similar spacetime is asymptotically flat when extended to
infinity (except flat spacetime); however this is no problem,
because the self-similar spacetime can be matched onto an
asymptotically flat spacetime at some large radius, and within this
radius the self-similar approximation is a good one; in fact, it
becomes increasingly good and natural as the collapse proceeds to
small scales at threshold
\cite{Chop,EC,HE}.

Section II reviews the calculation of critical exponents, while
Sects.~III and IV give the perturbation equations and their
boundary conditions.  Section V deals with gauge modes, while the
numerical methods are outlined in Sect.~VI.

Our main results are given in Sect.~VII\@.  Upon perturbing the
critical solution for the complex scalar field, we find two growing
modes.  The first mode is real; as expected it corresponds to
formation of black holes off threshold, and it yields a value $\gamma
\approx 0.387111\pm0.000003$.\@  The second growing mode --- actually
a conjugate pair of modes --- is oscillatory and must correspond to
an instability of the critical solution on threshold.

Finally Sect.~VIII discusses significance of the results.  We do
not attempt to follow the growth of the second growing mode outside
the range of validity of linear perturbation theory, but plausibly
it would develop into the Choptuik discretely self-similar
solution, the original choptuon \cite{Chop}.

\section{Critical Exponents and Perturbation Theory}

We briefly review the perturbation theory method
\cite{EC,KHA,Maison} and then apply it to our model.  Begin with a
continuously self-similar choptuon.  Perturb it.  Quite
generally, perturbed self-similar solutions to partial differential
equations show power-law behavior of modes\footnote{In degenerate
cases there may be logarithmic corrections to power-law behavior,
of the form $\epsilon (-t)^{-\kappa}\log^n (-t)$, but these turn
out to be absent in the problem at hand.} in time $t$, tantamount
to exponential behavior in logarithmic time $\tau\equiv\log(-t)$.
Consider a perturbation mode with power-law behavior $\epsilon
(-t)^{-\kappa}$, where $\epsilon$ $(\propto p-p^*)$ is a small
constant, and $t$ is a suitable time coordinate, in this case
proper time along the time axis $r=0$ before collapse at $t=0$.
Assume it develops into a small black hole.  The actual formation
of the black hole is outside the scope of perturbation theory, but
we can still determine the black hole mass by perturbation theory,
as follows.  By scale invariance of the choptuon, the formation of
black holes belonging to different values of $\epsilon$ will be
strictly homothetic to each other: {\it i.e.,} the full spacetimes
will be related to each other by exact scale transformations.
Determine the mass $M$ of the black hole as follows:  Pick a
fiducial amplitude of the perturbation mode, say small $\delta$,
which is still within the range of validity of perturbation theory.
Then imagine evolving further by an exact calculation --- but the
exact calculation need not be done, because it will just scale.
The black hole mass $M_{BH}$ will be proportional to the value
$t_1$ at which
\begin{equation}
 \epsilon (-t_1)^{-\kappa} = \delta\\
\end{equation}
or
\begin{equation}
  M_{BH} = {\rm const}(-t_1) =  {\rm const}'\epsilon^{1/\kappa}.
\end{equation}
The critical exponent, Eq.~(\ref{bhmass}), can then be read off as
\begin{equation}
	\gamma = 1/\kappa.
\label{getg}\end{equation}
To obtain the critical exponent $\gamma$, it thus suffices to find
the appropriate unstable mode in perturbation theory, and its
characteristic growth rate $\kappa$.

\section{Equations of Motion}

To perturb about a self-similar critical solution (choptuon), it is
convenient to adopt coordinates $\tau = \log(-t), z=-r/t$
where $r$ is the usual areal coordinate that measures the area of
2-spheres in spherical symmetry, and $t$ is an orthogonal time
coordinate, chosen to agree with proper time along the time axis
$(t<0,r=0)$.  With these coordinates, the time-dependent spherically
symmetric metric can be written,
\begin{equation}
	ds^2 = e^{2\tau}\left( (1+u)\left[-(b^2-z^2)d\tau^2 +
	2zd\tau dz+dz^2\right] + z^2d\Omega^2 \right).
\label{metric}\end{equation}
To enforce regularity along the time axis $z=0$ at the center of
spherical symmetry, the metric functions $u(\tau,z)$ and
$b(\tau,z)$ must obey boundary conditions
\begin{equation}
	b(\tau,0)=1,\quad u(\tau,0)=0
\label{bcmetric}\end{equation}

The perturbed fields are then defined as
\begin{mathletters}\label{pertvars}\begin{eqnarray}
  b(\tau,z)	&\approx& b_0(z) + \epsilon b_{1}(\tau,z) \\
  u(\tau,z)	&\approx& u_0(z) + \epsilon u_{1}(\tau,z) \\
  \phi(\tau,z)	&\approx& e^{i\omega\tau}\phi_0(z) + \epsilon
	e^{i\omega\tau}\phi_{1}(\tau,z)
\end{eqnarray}\end{mathletters}%
where ${}_0$ denotes the 0th order critical solution, ${}_1$
denotes the 1st order perturbation, $\omega=1.9154446$ is
the (unique) eigenvalue of the unperturbed equations (paper I); and
where $\epsilon>0$ is an infinitesimal constant, a measure of how
far away the solution is from the critical solution in the space of
initial conditions.  Using Choptuik's terminology, we consider the
supercritical regime for infinitesimal
\begin{equation}
	\epsilon \propto p - p^*.
\end{equation}

Matter consists of a free, massless,\footnote{In view of previous
results \cite{Chop} it is unlikely that a nonvanishing rest mass,
or more generally a nonlinear self-coupling of the form
$\Box\phi=V'(\phi)$ would make any difference below.  However, other
self-couplings such as in the axion-dilaton system \cite{EHH} may
well make a difference.} complex scalar field $\phi$ that obeys the
scalar wave equation
\begin{equation}
	\Box\phi = 0
\end{equation}
while the Einstein equations are
\begin{mathletters}\begin{eqnarray}
 R_{\mu\nu} - {1\over2}g_{\mu\nu}R	&=& 8\pi T_{\mu\nu}\\
	&=& 8\pi{\rm Re}\left(\nabla_\mu\phi \nabla_\nu\phi -
	{1\over2} g_{\mu\nu}\nabla^\rho\phi\nabla_\rho\phi\right);
\end{eqnarray}\end{mathletters}
paper I gives these equations in detail.

We now perturb these Einstein-scalar equations through 1st order in
$\epsilon$, to obtain a set of linear partial differential
equations for the perturbed fields $b_1$, $u_1$, $\phi_1$, in the
independent variables $\tau$, $z$.\@ Following the standard
approach, we Fourier transform the 1st order fields with respect to
the ignorable coordinate $\tau = \log(-t)$,
\begin{mathletters}\label{fourier}\begin{eqnarray}
  \hat{u}_1(\sigma,z)	&=& \int e^{i\sigma\tau} u_1(\tau,z)d\tau,\\
  \hat{b}_1(\sigma,z)	&=& \int e^{i\sigma\tau} b_1(\tau,z)d\tau,\\
 \hat\phi_1(\sigma,z) &=& \int e^{i\sigma\tau}\phi_1(\tau,z)d\tau;
\end{eqnarray}\end{mathletters}%
throughout, $\hat{}$
will denote such a Fourier transform.  The transform coordinate
$\sigma$ is in general complex.  The 1st order field equations now
become ordinary differential equations (ODE's) in $z$, and under
appropriate boundary conditions, become an eigenvalue problem for
$\sigma$\null.  Solutions of the eigenvalue problem are then normal
modes of the critical solution.  Eigenvalues in the lower half
plane ${\rm Im}\sigma<0$ belong to unstable (growing) normal modes.
Eigenvalues in the upper half $\sigma$ plane --- which we will not
explore in this paper --- would correspond to quasi-normal (dying)
modes of the critical solution.  The eigenvalue $\sigma$ is related
through Eq.~(\ref{getg}) to the critical exponent by $\gamma =
-1/\rm{Im}\sigma$.

Thereby, the Einstein-scalar equations reduce to the following set
of ODE's in $z$.  Define some auxiliary fields in 0th order
\begin{eqnarray}
q(z) &=&  \phi_0(z)'  \\
p(z) &=&  {1 \over b_0} ( i\omega \phi_0(z) - z \phi_0(z)') \\
\noalign{\hbox{and in 1st order}}
r(\sigma,z) &=&  \hat\phi_1(\sigma,z)  \\
s(\sigma,z) &=&  {1 \over b_0} \left(i(\omega-\sigma)\hat\phi_1 - z
   \hat\phi_1') \right)  \\
R(\sigma,z) &=&  r(-\sigma^{*},z)^{*} \\
S(\sigma,z) &=&  s(-\sigma^{*},z)^{*}.
\end{eqnarray}
where ${}^*$ denotes complex conjugate.
The field equations become (in notation which closely follows
that of paper I), in 0th order,
\begin{mathletters}\label{eqn0}\begin{eqnarray}
  b_0'	&=&  {b_0u_0 \over z}, \\
  u_0	&=&  4\pi z^{2}\left( |q|^{2} + |p|^{2} + 2{b_0 \over z}
	{\rm Re}(qp^{*}) \right),\\
  u_0'	&=& -4\pi b_0(1+u_0)(qp^*+pq^*), \\
  q'	&=& -\left( {u_0+2\over z} + {z\beta_+\over\Delta} \right)
        q + {b_0\beta_- \over \Delta}p,   \\
   p'	&=&   {b_0\beta_+ \over \Delta}q - \left( {u_0 \over z} +
        {z\beta_- \over \Delta} \right) p,
\end{eqnarray}\end{mathletters}%
and in 1st order
\begin{mathletters}\label{eqn1}\begin{eqnarray}
r' &=&   \left( {2 z b_0\hat{b}_1 \over \Delta^2} \beta_+ -
                 {\hat{u}_1 b_0^2 \over z\Delta} \right) q +
 \left( -{2 \hat{b}_1 b_0^2 \over \Delta^2} \beta_- + {i\sigma
 \hat{b}_1 + \hat{u}_1 b_0 \over \Delta} \right) p - \nonumber\\
&&\qquad \left( {u_0 + 2 \over z} + {z (\beta_+ - i\sigma) \over
  \Delta}\right) r + {b_0 \over \Delta}(\beta_- -i\sigma) s,\\
s' &=&   \left( -{2 \hat{b}_1 z^2 \over \Delta^2} \beta_+ +
 {\hat{u}_1 b_0 \over \Delta} \right) q + \left( {2 \hat{b}_1 b_0 z
 \over \Delta^2} \beta_- - {z\hat{u}_1 \over \Delta} - i\sigma {z
 \hat{b}_1 \over b_0 \Delta} \right) p + \nonumber\\
  &&\qquad {b_0 \over \Delta}(\beta_+ - i \sigma) r - \left( {u_0
  \over z} + {z (\beta_- - i\sigma) \over \Delta} \right) s,\\
R' &=&   \left( {2 z b_0\hat{b}_1 \over \Delta^2} \beta_+^{*} -
 {\hat{u}_1 b_0^2 \over z\Delta} \right) q^{*} + \left( -{2 \hat{b}_1
 b_0^2 \over \Delta^2} \beta_-^{*} + {i\sigma \hat{b}_1 + \hat{u}_1
 b_0 \over \Delta} \right) p^{*} -\nonumber\\
 &&\qquad \left( {u_0 + 2 \over z} + {z (\beta_+^{*} - i\sigma) \over
 \Delta}\right) R + {b_0 \over \Delta}(\beta_-^{*} - i\sigma) S,\\
S' &=&   \left( -{2 \hat{b}_1 z^2 \over \Delta^2} \beta_+^{*} +
 {\hat{u}_1 b_0 \over \Delta} \right) q^{*} + \left( {2 \hat{b}_1 b_0
 z \over \Delta^2} \beta_- - {z\hat{u}_1 \over \Delta} -
 i\sigma {z \hat{b}_1 \over b_0 \Delta} \right) p^{*} +\nonumber\\
  &&\qquad {b_0 \over \Delta}(\beta_+^{*} - I \sigma) R - \left({u_0
  \over z} + {z (\beta_-^{*} - i\sigma) \over \Delta} \right) S,\\
\hat{b}_1' &=&  {1 \over z} (\hat{u}_1 b_0 + u_0 \hat{b}_1),  \\
\hat{u}_1' &=&  -{\hat{u}_1 (1+u_0) \over z} + {u_0' \hat{u}_1 \over
  1 + u_0 } + 4\pi(1+u_0)z \left(-{2 \hat{b}_1 \over
  b_0} |p|^2 + p S + p^{*} s + q^{*} r + q R \right).
\end{eqnarray}\end{mathletters}%
Here
\begin{mathletters}\begin{eqnarray}
	{\hbox{\ }'}		&=& {d\over dz},\\
	\beta_{\mathord\pm}	&=& i\omega + u_0 \pm 1,\\
	\Delta			&=& b_0^2-z^2.
\end{eqnarray}\end{mathletters}%
The equations for $R$ and $S$ are identical to those for $r$ and
$s$ under the conjugacy $(q,p,r,s,\beta_{\pm})
\rightarrow (q^{*},p^{*},R,S,\beta_{\pm}^{*})$.  In the 0th order
problem, the Bianchi identities allow us to determine $u_0$ as an
algebraic expression, Eq.~(\ref{eqn0}b)\null.  Similarly
$u_{1}(z)$ can be written as a 1st order algebraic expression
involving other 1st order variables:
\begin{equation}
  \hat{u}_1 = {4\pi z (u_0+1) \over u_0+1 - i\sigma} \left( -{2zb_1
  \over b_0} |p|^{2} + z (qR + q^{*}r + pS + p^{*}s) +
           b_0 (pR + p^{*}r + qS + q^{*}s) \right)  .
\label{getu1}\end{equation}

\section{Boundary Conditions}

Now we specify boundary conditions for these equations.
On the axis of spherical symmetry, $z=0$, solutions must be regular.
This demand leads to boundary conditions in 0th order
\begin{mathletters}\label{bc0}\begin{eqnarray}
	b_0(0)	&=& 1, \\
	u_0(0)	&=& 0, \\
	q(0)	&=& 0, \\
	p(0)	&=& \hbox{free real const};\\
\end{eqnarray}\end{mathletters}%
and in 1st order
\begin{mathletters}\label{bc1}\begin{eqnarray}
	\hat b_1(0)	&=&  0,  \\
	\hat u_1(0)	&=&  0,  \\
	r(0)		&=&  0,  \\
	R(0)		&=&  0,  \\
	s(0)		&=& \hbox{free complex const}, \\
	S(0)		&=& \hbox{free complex const}.
\end{eqnarray}\end{mathletters}%
The global $U(1)$ phase symmetry present in the unperturbed
equations has allowed us to set $p(0)$ to be real.  No such
rotation can be performed for the time-dependent quantity
$\phi_1$, so $s(0)$ and $S(0)$ are in general complex.

The field equations have a regular singular point at the point
$z=z_2$ where $\Delta(\equiv b_0^2 - z^2)$ vanishes.  As discussed
in paper I, this locus corresponds to a similarity horizon in
spacetime, the past light cone of the spacetime singularity at the
spacetime origin $(t,r)=(0,0)$\@.  All fields must be regular on
this horizon, because it lies within the domain of dependence of
the initial hypersurface.  This means in particular that the terms
proportional to $1/\Delta$ in Eqs.~(\ref{eqn1}ab) must cancel at
$z_2$, giving a linear relation between $r(z_{2})$, $s(z_{2})$,
$\hat{b}_1(z_2)$ and $\hat{u}_1(z_2)$; and similarly for $R(z_{2})$
and $S(z_{2})$ from Eqs.~(\ref{eqn1}cd):
\begin{mathletters}\label{bcrs}\begin{eqnarray}
\lefteqn{(u_0-1)\left((\beta_+ - i\sigma) r - (\beta_- - i\sigma) s
	\right) = q \left({\hat{b}_1 \beta_+ \over z } +
   (i\omega + 2) \hat{u}_1 + \hat{b}_1 u_0' \right) - }\nonumber\\
  &&\quad p\left({u_0 \hat{b}_1(\beta_- - i \sigma)
   \over z} +{i\sigma \hat{b}_1 \over z} + \hat{b}_1 u_0' + i\omega
   \hat{u}_1 \right) + \hat{b}_1 (\beta_+ q' - \beta_- p')  \\
\lefteqn{(u_0-1)\left((\beta_+^{*}-i\sigma)R - (\beta_-^{*}-i\sigma)S
	\right) = q^{*}\left({\hat{b}_1 \beta_+^{*} \over z} +
  (-i\omega + 2) \hat{u}_1 + \hat{b}_1 u_0' \right) - }\nonumber\\
  &&\quad p^{*}\left({u_0 \hat{b}_1(\beta_-^{*} - i \sigma)
  \over z} +{i\sigma \hat{b}_1 \over z} + \hat{b}_1 u_0' - i\omega
   \hat{u}_1 \right)+\hat{b}_1 (\beta_+^{*} q^{*\prime{}} -
   \beta_-^{*} p^{*\prime{}}).
\end{eqnarray}\end{mathletters}%
These equations may conveniently be used to solve for $s(z_2)$ and
$S(z_2)$ in terms of other boundary conditions as long as
$\sigma$ lies in the lower half plane;  if one wished to explore
the upper half plane this procedure would fail somewhere due to
the vanishing of the coefficients of $s$ and $S$ at $z_2$.
The resulting boundary conditions at $z=z_2$ are as follows,
in 0th order:
\begin{mathletters}\label{bc2}\begin{eqnarray}
	b_0(z_2)	&=& z_2 = \hbox{free real const,}\\
%%	u_0(z_2)	&=& \hbox{free real const,}\\
	p(z_{2})	&=& \hbox{free complex const;}\\
\end{eqnarray}\end{mathletters}
and in 1st order:
\begin{mathletters}\label{bc3}\begin{eqnarray}
	\hat{b}_1(z_2)	&=& \hbox{free complex const,}\\
	\hat{u}_1(z_2)	&=& \hbox{free complex const,}\\
	r(z_2)		&=& \hbox{free complex const,}\\
	R(z_2)		&=& \hbox{free complex const.}
\end{eqnarray}\end{mathletters}

The 0th order equations, Eqs.~(\ref{eqn0}), amount to five real
ODE's in five real fields $(b,p,q)$, where the complex fields $p$
and $q$ each count as 2 real fields.  Correspondingly there are 5
free real constants: $(\omega,p(0),z_2,p(z_2))$, where the complex
$p(z_2)$ counts twice.\footnote{In paper I, we regarded $u$ as an
unknown rather than solving for it from Eq.~(\ref{eqn0}b);
therefore the counting was slightly different.}  Thus we have a
well posed eigenvalue problem on a finite domain $0\le z\le z_2$,
and we may expect a discrete spectrum.

The 1st order equations, Eqs.~(\ref{eqn1}) amount to 6 complex
linear ODE's\footnote{One could alternatively use the algebraic
constraint, Eq.~(\ref{getu1}), for $\hat{u}_1(z)$ to reduce the
number of complex equations to 5.  However, the denominator in
Eq.~(\ref{getu1}) vanishes on a region of the negative imaginary
$\sigma$ axis, and causes numerical trouble.} in the 6 complex
fields $(\hat{b}_1,\hat{u}_1,r,s, R,S)$\null.  At our disposal are
$(\sigma,s(0),S(0),\hat{b}_1(z_2),\hat{u}_1(z_2),r(z_2),R(z_2))$,
7 free complex constants.  Do we have one too many free complex
constants?  No, because solutions to the equations must come in
1-complex-dimensional linear families, due to linearity; that is,
each solution may be freely scaled by an arbitrary complex factor.
Thus we have a well posed eigenvalue problem on a finite domain
$0\le z\le z_2$ which should show a discrete spectrum of solutions,
apart from linearity.  We note that the 1st order system,
Eqs.~(\ref{eqn1}), with its boundary conditions, is not
self-adjoint, at least not obviously so; therefore the eigenvalue
spectrum of $\sigma$ may be expected to be complex.

Once constructed on the domain $0\le z\le z_2$, the 1st order
solution could be extended throughout the spacetime.  In the
terminology of paper I, Fig.~5, the 1st order solution is first
constructed in region I, and then it could be extended through
regions II and III\null.  We will have no need to carry out
that extension in this paper.

\section{Coordinate and Gauge Conditions}

Since our field equations possess gauge invariance due to general
coordinate invariance and global $U(1)$ phase invariance, some
unphysical pure gauge modes will appear at 1st order, to the extent
that the gauge conditions implicit in our boundary conditions
Eqs.~(\ref{bc0}) fail to be unique.

Indeed there arises a pure gauge mode from an infinitesimal phase
rotation $\phi\rightarrow e^{i\epsilon}\phi$ in the 0th order
critical solution:
\begin{mathletters}\label{gauge1}\begin{eqnarray}
	\hat{b}_1(z)	&=& 0,\\
	\hat{u}_1(z)	&=& 0,\\
    \hat\phi_1(z)	&=& i\phi_0(z).
\end{eqnarray}\end{mathletters}%
This gives a time independent solution of Eqs.~(\ref{eqn1}) that
satisfies the boundary conditions; hence it corresponds to an
unphysical mode at $\sigma=0$

A second pure gauge mode arises upon adding an infinitesimal
constant to time $t\rightarrow t+\epsilon$ at constant $r$ in the
0th order critical solution.  This is possible because our
coordinate conditions, Eqs.~(\ref{bcmetric}) normalize $t$ to
proper time along the negative time axis $(t<0,z=0)$, but the zero
of time is not specified.  Then the solution is perturbed by
\begin{mathletters}\label{gauge2}\begin{eqnarray}
    b_1(\tau,z)	&= {\partial b_0\over\partial t}|_r &= -(z/t)b'(z) =
				e^{-\tau}zb'(z),\\
    u_1(\tau,z)	&= {\partial u_0\over\partial t}|_r &= -(z/t)u'(z) =
				e^{-\tau}zu'(z),\\
 \phi_1(\tau,z)	&= e^{-i\omega\tau}{\partial(e^{i\omega\tau}\phi_0)\over
	\partial t}|_r &= e^{-\tau}(-i\omega\phi_0(z)+z\phi_0'(z)).
\end{eqnarray}\end{mathletters}%
This pure gauge mode has time dependence $e^{-i\sigma\tau} =
e^{-\tau}$ and so has negative imaginary $\sigma=-i$.  There are no
other pure gauge modes in the $\sigma$ plane.  These two modes
should appear as numerical solutions, but are unphysical.

\section{Numerical Method}

To solve the 1st order problem we used a Runge-Kutta integrator with
adaptive stepsize as part of a standard two point shooting
method \cite{NR}, shooting from $z=0$ and from $z=z_2=5.0035380$, and
matching in the middle $z=z_{m}=2.5$.  For convenience we solved
the 0th order system, Eqs.~(\ref{eqn0}), and the 1st order system,
Eqs.~(\ref{eqn1}) simultaneously with the same steps in $z$.  As
discussed in paper I, the similarity horizon $z_2$ is a demanding
place to enforce a boundary condition, and a second order Taylor
expansion of the regular solution was used for this purpose.

To solve the 1st order system, we collected all the boundary values
but $\sigma$ into a complex 6-vector
$X\equiv(s(0),S(0),\hat{b}_1(z_2),\hat{u}_1(z_2),r(z_2),R(z_2))$.
Because the equations are linear, the matching conditions at
$z=z_{m}$ are likewise linear in the boundary values.  A solution
is found when the values at $z_m$ of $(\hat{b}_1,\hat{u}_1,r,s,R,S)$
upon integrating from $z=0$ match with those found by integrating
from $z=z_2$, for some boundary values $X$.  We can express this
matching condition
\begin{equation}
	A(\sigma) X = 0
\end{equation}
where $A(\sigma)$ is a $6\times6$ complex matrix which is a
nonlinear function of $\sigma$, constructed numerically by
integrations of the 1st order equations, Eqs.~(\ref{eqn1}), for six
linearly independent choices of boundary values $X$.  The condition
on $\sigma$ for a solution is then
\begin{equation}
	\det A(\sigma)=0.
\end{equation}
Once a value for $\sigma$ was found that satisfies this condition,
the corresponding boundary values $X$ were found as a zero
eigenvector of the matrix $A$; these come in one (complex)
parameter families, as observed above.  Solution of
Eqs.~(\ref{eqn1}) with boundary values $X$ yields the normal mode
itself.

Now, $A(\sigma)$ has been carefully constructed so that it is a
complex {\it analytic} solution of $\sigma$.  This follows from the
fact that all equations leading to $A$ contain $\sigma$ but not
$\sigma^*$, together with some standard theorems about ODE's.
Moreover, $A(\sigma)$ has no singularities in the closed lower half
$\sigma$ plane.  These properties allow us to use a number of ideas
from scattering theory to study $\det A(\sigma)$.  In particular,
there is a theorem for counting the number $N_C$ of zeros of $\det
A$ within any closed contour $\cal C$ in the closed lower half
$\sigma$ plane:
\begin{equation}
	\Delta_{\cal C}{\rm Arg}\det A = 2\pi N_{\cal C}
\label{count}\end{equation}
where ${\rm Arg}\det A$ is the phase of $\det A$, and $\Delta_{\cal
C}{\rm Arg}\det A$ is the total phase wrap (in radians) around the
closed contour ${\cal C}$, a result similar to Levinson's
theorem for counting resonances in quantum scattering theory.

Furthermore, there holds a conjugacy relation,
\begin{equation}
	A^*(-\sigma^*) = A(\sigma),
\end{equation}
which means that $A$ need only be evaluated for ${\rm Re}\sigma\ge0$
in the lower half plane.

The nonlinear equation $\det A(\sigma)=0$ was solved by the secant
variant of Newton's method \cite{NR}.  The equation being
complex-analytic, the 1-complex-dimensional realization of the
method was used, and it performed well.

The following numerical checks are used in our work.  First, the
two pure gauge modes, Eqs.~(\ref{gauge1}, \ref{gauge2}), appeared
numerically at the expected places $\sigma=0,-i$.  Second, the
constraint, Eq.~(\ref{getu1}), was confirmed numerically.  Third,
we numerically spot-checked the Cauchy-Riemann equations for $\det
A(\sigma)$, to confirm that this function is analytic.  Generally
the accuracy was about $10^{-6}$.

\section{Results}

There are five values of $\sigma$ in the lower half plane at which
$\det A$ vanishes, signalling five modes.  All five are simple
zeros; see Fig.~1\@.
\global\firstfigfalse
\begin{figure}
\null\vskip-1in
\centerline{\epsfysize=6in\epsfbox{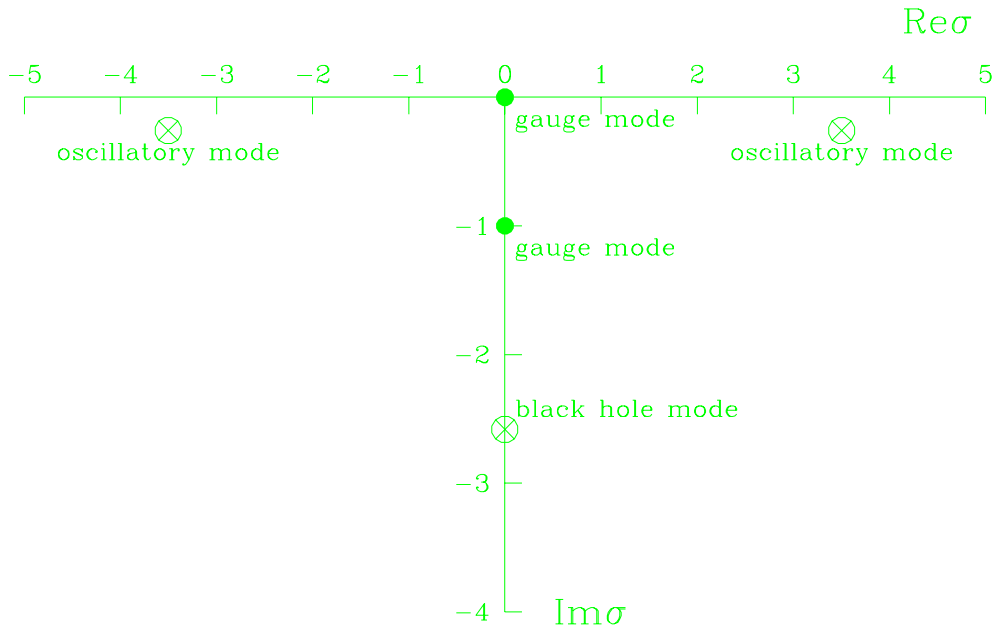}}
\null\vskip-1in
\caption{The lower half complex $\sigma$ plane.  There are five
zeros of $A(\sigma)$, representing five modes (solutions of the 1st
order perturbed field equations).
$\bullet$ $\bullet$ Two unphysical gauge modes.
$\otimes$ $\otimes$ $\otimes$ Black hole mode,
and pair of oscillatory unstable modes.
See Sect.~VII.}
\label{fig1}
\end{figure}
\vskip0.5in
Two of these modes are pure gauge modes, Eqs.~(\ref{gauge1},
\ref{gauge2}), which help check internal consistency of the work, but
are unphysical; as expected these lie on the negative imaginary
axis\footnote{To within ${\rm Re}\sigma=\pm10^{-6}$.} at
$\sigma=0,-i$.

The third solution also lies on the negative imaginary axis, at
\begin{equation}
	\sigma =-2.58324i \pm 0.00002i .
\label{mode3}\end{equation} We interpret this mode as the instability
toward forming an infinitesimal\footnote{In classical gravity.} black
hole, Eq.~(\ref{bhmass}).  This corresponds through Eq.~(\ref{getg})
to a critical exponent of $\gamma=0.387111\pm0.000003$.  This value
differs from that found by Choptuik \cite{Chop} for the real scalar
field, further evidence of non-universality.

The fourth and fifth solutions form a conjugate pair
lying at
\begin{equation}
	\sigma=\pm(3.50224 \pm 0.00001) - (0.25867 \pm 0.00001)i.
\label{mode4}\end{equation}
The presence of this pair of zeros in the lower half plane means
that the continuously self-similar critical solution is unstable to
perturbations, even among solutions ``tuned" to the black hole
threshold.  These represent a pair of modes that are not just
growing, but oscillating as well.  These modes must grow until they
become nonlinear, and then plausibly they go over to Choptuik's
solution.

We evaluated $\det A(\sigma)$ around a large rectangular contour
$\cal C$ bounded by $0\le{\rm Re}\sigma\le25$, $0\ge{\rm
Im}\sigma\ge-10$ (detouring around the three zeros on the imaginary
axis) and used Eq.~(\ref{count}) to count the zeros lying within.
The absence of other zeros in this large rectangle was thereby
established.  Asymptotically, we observed
\begin{equation}
	\det A(\sigma)\sim\exp(\tau_\infty\sigma),
	\quad\tau_\infty\approx6i
\end{equation}
in both the real and imaginary directions, and the outer boundaries
of the large rectangle appeared to be well into the asymptotic
regime.  Thus it appears very likely that there are no zeros in the
lower half $\sigma$ plane other than the five described above.

A comment is in order concerning the error of our numerical solution.
The error for the complex Choptuon constructed in paper I was a
little better than $10^{-6}$.  To investigate how the error of the
unperturbed solution affected the results in the perturbed problem, we
varied the parameters of the unperturbed Choptuon by $10^{-5}$.\@  The
effect on the above zeros was a variation of about $10^{-5}$.\@  We
also varied the initial values for the perturbed problem
(independent of the change in the unperturbed choptuon) in order to
study whether the locations of the zeros were sensitive to this
variation.  The resulting change in the values was on the order of
$10^{-6}$.  This exercise also verified that the Taylor expansions at
the endpoints 0 and $z_2$ are of adequate order.

Since our integrator uses an adaptive stepsize, we did not vary the
stepsize directly;  rather we varied the algorithmic error tolerance
used to adjust the stepsize, and we allowed the initial stepsize,
$\epsilon$, and the location of the matching point, $z_{m}$, to be
random numbers within a specific range:
\begin{equation}
10^{???} < \epsilon < 10^{-6},  \quad 0.5 < z_{m} < 4.5\,.
\end{equation}

The combined effect of varying the parameters of the original
choptuon by several times their quoted error, and of varying the
initial values and other parameters for the perturbed problem, then
gave the overall uncertainties quoted above.

Thus we conclude that the critical exponent obtained here,
$\gamma=0.387111\pm0.000003$, is indeed different from others that
have been calculated to date, such as $\gamma_{\rm rad} = 0.3558019$,
as found for radiation fluid collapse by Koike, Hara \& Adachi
\cite{KHA}.

\section{Discussion and Conclusion}

We have perturbed the continuously self-similar complex choptuon
\cite{HE} and find the critical exponent for black hole mass,
Eq.~(\ref{bhmass}) to be
\begin{equation}
	\gamma=0.387111\pm0.000003,
\end{equation}
further evidence for non-universality of $\gamma$.  We also
find a further, oscillatory instability in this choptuon.

In spherically symmetric collapse of a real scalar field, Choptuik
\cite{Chop} observed that the real, discretely self-similar choptuon
could be constructed by tuning only one (real) parameter $p$ of the
infinite number of parameters in the space $\cal S$ of initial
conditions.  That is, the real choptuon is an attractor in a
subspace of codimension 1 in $\cal S_{\hbox{real}}$.  The results
here for complex scalar fields imply that the continuously
self-similar complex choptuon would require tuning of some three
(real) parameters $p$, $q$, $r$ in the space $\cal
S_{\hbox{complex}}$ of spherically symmetric initial conditions ---
one ($p$) to control the real black hole mode, two more ($q+ir$) to
control the pair of oscillatory modes.  Therefore the complex
choptuon appears to be an attractor of codimension 3 in $\cal
S_{\hbox{complex}}$  On threshold --- with only $p$ tuned --- it is
highly plausible that the oscillatory instability found here in the
complex choptuon would evolve nonlinearly toward the original real
choptuon.  Otherwise, it would have to evolve toward still a third
choptuon, for which there is currently no evidence.

Numerical relativity \cite{Chop,CPP,DG} is needed to follow this
nonlinear evolution.  In fact, the growth rate of the oscillatory
mode (measured by $-{\rm Im}\sigma$) is small, $\sim10$\% of the
growth rate of the black hole mode, Eqs.~(\ref{mode3},
\ref{mode4}), and consequently the oscillatory instability may be
hard to isolate numerically.  It would be interesting to study
related models such as a complex scalar field with a conformal
coupling $\xi$, or the axion/dilaton fields \cite{EHH}, to see what
happens to the oscillatory growing mode.  This mode might
conceivably move into the upper half $\sigma$ plane to become
stable.

Choptuons thus show both similarities and differences compared to
stationary black holes.  For classical black holes, the uniqueness
theorems show that ``black holes have no hair'', and perturbation
theory proves that black holes are stable in vacuum, or when
coupled to massless fields.  On the other hand, classical choptuons
coupled to a massless complex scalar field show at least two
alternative states, namely the original choptuon \cite{Chop} and
the complex choptuon \cite{HE}\null.  In this paper, perturbation
theory has shown that the latter state is unstable, presumably
toward the former.

\acknowledgements

This research was supported in part by the National Science
Foundation under Grant Nos.~PHY89-04035 and PHY90-08502, and parts
were carried out at the Aspen Center for Physics.  We are grateful
to Matt Choptuik, Chuck Evans and Jim Horne for enlightening
conversations and communications.

\end{document}